\documentstyle[11pt,newpasp,twoside]{article}
\def\edcomment#1{\iffalse\marginpar{\raggedright\sl#1\/}\else\relax\fi}
\reversemarginpar

\begin{document}
\title{Energy Release and Transport Processes in the Centres of Galaxies}
 \author{Roger Blandford}
\affil{130-33 Caltech, Pasadena, CA 91125, USA}
\begin{abstract}
Observations over the past decade have verified, beyond reasonable doubt,
that most galactic nuclei contain massive black holes. Hole masses are  being 
measured and firm evidence for spin is being sought.
Attention is now returning to the study of how gas flows around black holes 
and how energy is released both from the accreting gas and from the hole
itself in the form of radiation, relativistic jets and non-relativistic, hydromagnetic
winds. Some of the different possibilities 
currently under investigation are briefly reviewed
and some recent clues from radio scintillation, polarization and X-ray observations
are discussed. It is argued that observations of persistent
circular polarisation in Sgr A$^\ast$ support the presence of an ordered disk magnetic
field. It is also conjectured that adiabatic, 
sub- and super-critical accretion flows are 
demand-limited, not supply-driven and are associated with large mass
outflow as appears to be the case in Sgr A$^\ast$. This principle 
may have to be modified when a massive black hole forms in
a protogalaxy and this modification may account for
the proposed hole mass-bulge velocity dispersion relation.   
The final stages of this process may release sufficient
wind energy from the nucleus to prevent disk formation.
\end{abstract}
\section{Introduction}
This is an exciting time in the study of active galactic nuclei. Radio, optical and X-ray
observations are not only persuading the last doubters that most normal galaxies contain
massive black holes, but are measuring of their masses [Dressel, Baan, Greenhill] and
probing their spins. Furthermore, the high 
angular resolution accessible to ground-based and space-based
observation of relativistic jets is allowing us to trace these outflows down to radii
less than  $\sim100m$ vindicating the view that they 
are at least powered by physical processes
located fairly close to the central black hole
[Falcke, Hirabayashi]. Finally, radio polarimetric and $\gamma$-ray [Edwards] 
observations of these relativistic jets are being interpreted in a manner that allows us 
to make arguments concerning their composition, which, in turn, impact our understanding
of how they are formed. 
Many of these topics are described at greater length elsewhere in these proceedings
and designated with square brackets.
I shall just emphasize some more recent theoretical ideas.
\section{Accretion Power vs Black Hole Power}
As is often the case, the most important factor controlling the evolution
of accretion disks is the flow of energy. There are two power
sources. Firstly, accreting gas will 
release its binding energy before it makes its final adiabatic plunge towards the event 
horizon. However, the
energy will generally not be liberated all at the point of release
because the 
magnetic torques which transport angular momentum, also transport energy outward
radially through the disk.
(There can be additional, convective and radiative energy transport which 
is ineffective in transporting angular momentum.) 

There are several possible channels through which this energy can escape and their 
relative importance surely depends upon the accretion rate. It can be radiated 
from a disk photosphere. It can emerge from the disk surface as a hydromagnetic 
energy flux and heat an active corona, which can in turn radiate non-thermally
(for example by Comptonising disk photons) or drive a wind. The coronal heating
will probably occur via magnetic reconnection or a turbulent cascade. 
Energy and angular momentum can also be extracted from the disk if there
is a large scale field that launches a hydromagnetic wind.
There will then be an external torque per unit radius $dG/dr$ which will
extract energy at a rate $\int dG\Omega$ where $\Omega$ is the
disk angular velocity without dissipation. 

The second source of energy is the spin of the hole. As is well known, up to 29 percent 
of the mass of a spinning black hole is, in principle, 
extractable (eg Lee, Wijers \& Brown 1999). 
This is $6\times10^{62}(M/10^9{\rm M}_\odot)$~erg. In practice no more than,
say, a quarter of this energy is likely to be available, but still it is ample to account
for the non-thermal emission associated with the most profligate quasars. It 
is important to realize that, although energy is not strictly localisable 
in general relativity,
the spin energy is just as much associated with the so-called ergosphere
outside the event horizon as with the interior of the hole.  (The ergosphere
can be defined as the region within which a distant observer must rotate with respect
to the non-rotating frame at infinity.) This is a crucial and important difference 
from a spinning, newtonian star. 

Again there are many possibilities [Meier].
It is possible to extract work from orbiting matter which can be dragged by 
magnetic stress 
on to lower or even negative energy (including rest mass) orbits within the ergosphere
(e. g. Ruffini \& Wilson 1975; Agol \& Krolik 2000). When this 
matter enters the horizon, the net gravitational mass of the hole will
increase by less than the accreted rest mass and will actually decrease
if the orbit has negative energy. If the matter is attached to open magnetic field lines,
then work is done on the base of the jet and drives an outflow, 
either continuously or intermittently. Alternatively, if the infalling matter
is magnetically attached to the disk and still in causal contact, that is to say outside
the relevant, ingoing critical points, then orbital energy and angular 
momentum can be transferred to the disk from within its innermost stable circular 
orbit, thereby enhancing the disk power. In either case, it is an arbitrary choice
whether to charge the energy extracted to the hole or the accreting matter. 
Magnetic flux that threads the horizon can transfer energy directly from the hole
to the jet or the disk
at a rate $\sim a^2B^2c$, where $a$ is the specific angular momentum of the hole
and $B$ is the field strength. To an observer hovering
just above the horizon and, necessarily, orbiting with the hole angular velocity, the 
electromagnetic energy flux
will appear to enter the hole. However energy flux is conserved in the non-rotating frame
and transforming into this frame produces an outward energy flux at the
horizon. In other words,
the black hole, a spinning conductor, acts like a giant battery. Under the conditions
invoked for Cygnus A, say, an EMF of $\sim10^{20}$~V is 
induced, with a field $B\sim10^4$~G,
and the current is $\sim10^{18}$~A. (The resistance $\sim100\Omega$ turns
out to be similar to the impedance of free space.
The merit of this process as a candidate for powering
relativistic jets is that the electromagnetic power is likely to be relatively 
uncontaminated with baryons, the disadvantage is that if the field strength above the hole
is comparable with that above the inner disk then, the relatively small 
area of the horizon
will ensure that the hole power be less than the disk power, which, although
adequate for most jets may not be sufficient for the most luminous examples. 
Note, though, that this conclusion
does not follow if, as commonly envisaged, the disks associated with powerful radio
galaxies are thick. 

There is actually a second way to extract energy from the hole. 
If the disk plane wobbles about the equatorial plane of the hole, then there 
will be a ``Bardeen-Petterson'' torque that acts between the disk and the hole.
The nature of this torque has been quite controversial over the years, but the
net result is that the disk will be brought into alignment with the hole within 
some radius, $r_{{\rm BP}}$. There will be a reaction torque operating on the hole
and an associated dissipation in the disk near $r_{{\rm BP}}$. 
This could provide a significant
local heating of the disk under some models (eg Natarajan \& Pringle 1998) 
\section{Jets}
So far, I have emphasized the (electro)magnetic extraction of energy from the
disk and the hole because most models are now of this type. However, this 
does not imply that observed jets mostly carry Poynting flux. Just as is the case with 
the Crab Nebula, there is believed to be a transformation of the energy flux
from mechanical to electromagnetic to particle. The issue now is where and how do these 
transitions occur. One extreme possibility is that a hot plasma forms through
magnetic dissipation in the corona itself and that this flows hydrodynamically 
into the jet. (This would probably have to be an electron-ion plasmas as
it is very hard to see how a pair plasma could be accelerated
to ultrarelativistic bulk speed close to the hole without suffering catastrophic radiative
drag.)  Another possibility is that ultrarelativistic ions are formed close
to the hole and somehow directed along a jet and that these cascade down to lower energy
particles that will produce the radiation we observe.  In many regards the most natural
scheme is that the energy is carried away electromagnetically from the vicinity 
of the hole and that the pair fraction gradually builds up at a rate that may 
be limited by annihilation. Provided that this happens after the radiative drag
is ignorable, an ultrarelativistic jet will ensue. 
  
Extracting the energy is only half of the challenge in 
forming a jet. The outflow must also
be tightly collimated. For a long while it has been thought that jets required disks.
Howevever, with the discovery of jets associated with the Crab and Vela pulsars, perhaps 
this is not even necessary.  Magnetic collimation is most commonly invoked simply because
it is hard to sustain the necessary gas pressure against large radiative loss
at the location where the collimation must be occuring. 
(Note that this does not imply that
gas pressure is irrelevant. Magnetic stress must be 
ultimately confined transversely at large
cylindrical radius and static gas pressure or ram pressure is surely responsible.) 

There are still choices that have to be made in magnetic collimation. 
The confining stress 
can be attributed to the pressure of poloidal magnetic field or the tension associated
with the toroidal field. Alternatively, the field may be quite disorganised but
show an anisotropic stress tensor which somehow achieves the necessary collimation. 

One argument that has often been made against magnetic collimation, particularly by
toroidal field is that such configurations are probably unstable. However, this may not
be such an unwelcome conclusion because observed jets often do appear to corkscrew in 
real space and this is a natural instability mode for a pinched jet. At an even more 
extreme level, many jets appear to be episodic and to exhibit
strong internal shocks [Klare]. This, too, can be regarded as the nonlinear development 
of a jet instability. One possibility is that the natural increase in the ratio 
of the toroidal to the poloidal field with distance along the jet is balanced
by steady magnetic reconnection, which limits the growth of toroidal field and lead to
the steady acceleration of electrons and positrons.

In order to construct more quantitative models of jets, it is necessary to make better 
estimates of the power that they carry (as distinct from the power that they radiate).
There are several approaches that have been followed. One is to assume that the pressure
is well-estimated by the equipartition pressure and that the jet expands on the Mach cone.
In this case the power will be $L_{{\rm j}}\sim P_{{\rm eq}} r^2 V_{{\rm j}}$, a formula 
that is roughly valid relativistically. Alternatively, if the external pressure is known 
and the jet is thought to be in pressure equilibriium with its surroundings, then we can 
replace the equipartition pressure with the external pressure. 
A third approach is to estimate
the power supply to large double lobes. Using arguments like this it has been 
possible to argue that the jet power in some sources, including M87 [Junor]
and some blazars, exceeds the bolometric
power. If this were generally true, 
then it would effectively rule out many jet formation models.
The mid infrared observations of SIRTF will be crucial for measuring the 
total nuclear luminosity of radio galaxies.
\section{Brightness Temperature and Polarisation}
One of the more interesting and potentially 
significant developments in recent years has been 
the accumulation of evidence for high brightness 
temperatures in compact extragalactic radio 
sources. Intraday variable sources exhibit 
measured brightness temperatures that are at least as high as $\sim3\times10^{13}$~K
and could be significantly larger [Jauncey, Dennett-Thorpe, Kedziora-Chudczer]. 
This is conventionally interpreted in terms of the
traditional inverse Compton limit. For the observed ratios of Compton to synchrotron power
in jets, the comoving brightness temperature has an upper limit of $\sim10^{12}$~K. One 
power of the Doppler factor, which is twice the relevant Lorentz factor converts to the 
comoving frame suggesting that $\Gamma>20$. (Large Lorentz factors are not 
so intimidating now as values of $\Gamma\sim300$ are derived 
for $\gamma$-ray bursts although the radiative efficiency may be low at
radio wavelengths.) If it turns out that comoving brightness temperatures well
in excess of the Compton limit are required then it may be necessary to invoke 
a coherent emission mechanism. This is not, by itself, unreasonable; after all,
the Sun and Jupiter support coherent emission under relatively benign conditions.
A cyclotron maser is a particularly enticing possibility.
What may be a bit more problematic is the escape of very high brightness radiation from the source
when it is subject to nonlinear induced Compton and Raman scattering
(Levinson \& Blandford 1995).  

Now, in a self-absorbed synchrotron source, close to the inverse Compton limit,
the energies of the emitting electrons at the radio photosphere must be $\sim300$~MeV.
We do not see the emission from lower energy electrons 
and it is interesting to ask if they 
are present and if the electrons that we do see have proton or 
positron partners. As argued by Wardle et al. (1998 and references
cited therein), in the former case, the lower cut off in the electron distribution
function must be $\sim50$~MeV, otherwise there will be too much jet kinetic energy,
carried mainly by the protons. In the latter case, the pair distribution function
will extend down to mildly relativistic energy  
Similar conclusions are drawn from the observation of strong linear polarization 
which limits the internal Faraday rotation within the source. Finally the detection 
of persistent circular polarisation [Macquart]
has been used to rule in favor of a pair plasma,
on the grounds that the frequency dependence is better fit by Faraday conversion
in a pair plasma than intrinsic circular polarisation in a proton plasma.
Note that if the radiation mechanism is coherent cyclotron radiation, then the fact that 
the polarisation is not even stronger may be an indication 
that there is a pair plasma present.
\section{Galactic Centre}
The 2.6 million solar mass black hole at our Galactic center is in many respects the best 
example to study because it is so inactive [Eckart]. It has 
recently been detected in X-rays
with a very soft
2--10~keV luminosity (Baganoff 2000) of 
$\sim2\times10^{33}$~erg s$^{-1}$, which may be variable. This is a strikingly 
low luminosity. The independently estimated mass supply rate is very roughly 
$\sim10^{22}$~g s$^{-1}$ and yet the bolometric luminosity is $\sim10^{36}$~erg s$^{-1}$
corresponding to an apparent radiative efficiency $\sim10^{-7}c^2$ [Coker]. 
As has been discussed
elsewhere by Blandford \& Begelman (1999), this is consistent with the 
theoretical prediction that adiabatic (in the sense that there is no loss of heat)
accretion flows lose a large fraction of their mass before crossing the 
horizon. This is in contrast to the prediction of the widely-discussed and 
conservative ADAF models. (A naive interpretation
of the X-ray observations suggests a maximum mass accretion rate of $\sim10^{20}
(T_e/10^{10}{\rm K})^{1/4}$~g s$^{-1}$. It could plausibly be much
lower.) The reported spectrum suggests that the X-ray 
emission is nonthermal, which is really expected, because it would be remarkable
if trans-sonic, trans-Alfv\'enic, trans-relativistic flows avoided
accelerating relativistic particles.

If the mass accretion rate really is very low, then this opens up the possibility 
that we may be seeing radio or, more likely, mm emission from very close to the hole
and that polarization observations might be quite diagnostic of the physical conditions.
Now, in the case of Sgr A$^\ast$, there is no linear polarization reported at radio
wavelengths (although it has been reported, quite surprisingly, at submm 
wavelengths). The lack of linear polarization is quite consistent with the high Faraday
rotation expected from an accretion flow onto a black hole.
However, variable circular polarization has been reported with a persistent sign.
There is the intriguing possibility that this reflects the orbital motion of the 
plasma around the central black hole. To see how this might come about,
consider a magnetoactive plasma with $X=\omega_P^2/\omega^2,Y=\omega_G/\omega<1$.
Under quasi-longitudinal conditions, essentially $\cos\theta<Y$, 
where $\theta$ is the angle between $\vec k$ and $\vec B$, the electromagnetic eigenmodes 
are elliptically polarized with axis ratio $r=1\pm Y\sin\theta\tan\theta$ 
and phase velocity difference $\Delta V=cXY\cos\theta$. 

Now consider synchrotron or cyclotron radiation emitted from within an accretion disk
of thickness $H$. The major axis of the polarization ellipse will be Faraday  
rotated at a rate $\Delta\Phi/ds=\Delta V\omega/2c^2$. Now, we expect
that the magnetic field will reverse often along a ray and as
$X|Y|\omega H/c$ is anticipated to be much greater than unity, then we expect that the 
emergent linear polarisation will be vanishingly small. (The limiting polarisation 
along an individual ray is likely to be determined by the decrease in the density rather 
than the magnetic field strength.) 

The circular polarisation is a bit more problematic. If we suppose that there is a net 
magnetic field normal to the disk, as is true of some models, then there should be a 
preferred sense of the circular polarization that is emitted in cyclotron
or low energy synchrotron radiation.  This will be largely
preserved in propagating out of the disk. 
However, eventually, the sign of the net vertical
field must change and we might even be able to see this change propagate from low 
to high frequency if, as seems reasonable, there is a radius to frequency mapping
in the disk. However, it is also possible that there is essentially no large scale field.
In this case, there is a second mechanism which may create circular polarisation.

Consider a single eigenmode propagating out of the disk through a spatially 
varying magnetic field. We suppose that the variation happens relatively slowly on the
scale of the wavelength so that the polarization adjusts adiabatically (with no
mode crossings). We ignore refraction and recall that the electric vector is normal
to $\vec k$. Next, suppose that the two eigenmodes are launched with equal amplitude and 
that the field is uniform. The beating between the two eigenmodes will result in a
circular polarisation of amplitude $Y\sin\theta\tan\theta$ that changes sign as the plane 
of linear polarisation rotates. The limiting circular polarisation that 
emerges from the disk is equally likely to have either sign and so there will 
be no net circular polarisation. Now let the magnetic field direction
vary along a ray. 
If the angle $\theta$ varies, then there will be a corresponding change in the
axis ratio of the polarization ellipse, but still no preference
for one sign over the other. However, if the azimuthal angle 
$\phi$ relating $\vec B$ to $\vec k$ changes in a systematic fashion along all rays then it appears that
the circular polarisation will have one sign
as opposed to the other for slightly more than half the time.
The average circular polarization along a ray will 
be $\sim<Y\sin\theta\tan\theta d\phi/d\Phi>
\sim|Y\ln Y|d\phi/d\Phi$. 

Now this is just what we might anticipate
in a magnetised accretion disk. In the disk interior, the typical field direction will 
trail to reflect the differential rotation in the disk. However, 
the field will be swept back by progressively smaller angles as the ray 
approaches the disk surface, corresponding to a net rotation of the average
azimuthal angle $\phi$. There will only be a preferred
sense of limiting circular polarization, if the magnetostatic field is still changing in
this systematic manner over the last radian of Faraday rotation. The net circular
polarization will be $\sim c\ln Y/\omega HX$, where $H$ is the disk thickness. However, 
if $X>10^{-6}$, as seems assured in 
Sgr A$^\ast$, then this mechanism is unlikely to create
the observed circular polarisation at radio 
wavelengths, although it may be relevant in the sub mm.
I conclude that the measurement of persistent circular polarization
is more naturally interpreted in terms of a long term, ordered 
magnetic field, as discussed above.
There may be a related effect that can occur in the ergosphere, where the dragging of inertial
frames  is reponsible for a rotation of the 
polarisation ellipse. These ideas are currently under study.
\section{Black Hole Formation}
The foregoing considerations apply to sub-critical accretion. Similar principles are
relevant to super-critical accretion. Here the problem is not creating the 
photons, but allowing them to escape. Electron scattering opacity almost certainly 
dominates the radiative transfer and the photons will be trapped out to a radius
$r_{{\rm tr}}=\dot M\kappa_T/4\pi c$. Within this radius, we can regard the inflow as 
adiabatic, with an outward transport of energy and angular momentum. 
I have conjectured elsewhere that supercritical
accretion is ``demand-limited'' rather than ``supply-driven''. In other words 
black holes only grow at roughly the critical rate, 
$\dot M=4\pi GM/\epsilon\kappa_Tc$, where $\epsilon\sim0.1$ is the 
overall efficiency of energy release. If a black hole is supplied at a greater
rate, it will radiate at a few times the Eddington luminosity and the excess mass
and energy will be carried off in an outflow with a similar power. The mass 
may be liberated at all radii so that
the local net mass accretion rate grows roughly linearly with radius.
Alternatively, there could be strong, internal circulation
which carries the energy released near the black hole  
to some outer radius from where the wind is launched. 
This behaviour is, arguably, observed in Broad Absorption Line Quasars
and some of the most luminous, Galactic, transient sources [Spencer].

If this principle were correct, it would raise an important 
question. How do black holes grow to a billion solar masses,
when the universe was less than $\sim1$~Gyr old (as observations
of high redshift quasars now suggest) if there were only time for a 
few e-foldings in mass? One possible answer is that 
large holes might have formed by dynamical collapse. However, this 
seems a bit unlikely as they should have made 
a sort of mega gamma ray burst, and there is no evidence for such
prolonged, luminous explosions on our past light cone. 
Alternatively, what may be different about accretion during 
the early epochs of galaxy formation
is that the black hole should be surrounded by a chaotic, 
self-gravitating reservoir of gas, extending from 
the Bondi radius, $r_{{\rm B}}\sim GM/\sigma^2$ where $\sigma$
is the velocity dispersion determined by the dark matter,
stars and gas to some outer radius, $r_{{\rm gal}}$.
We expect the gas initially to dominate the mass in 
the inner parts of the galaxy.

Now, modifying a suggestion of Silk \& Rees (1998),
suppose that a small black hole forms. The maximum, net mass supply rate
is the Bondi (or Jeans') rate, $\dot M\sim\sigma^3/G$ under 
these conditions. If this were
to accrete onto the black hole and release energy efficiently,
the resulting wind would quickly blow away all the surrounding gas and shut off the 
supply. What seems most likely is that the feedback limits the accretion rate
to a value such that the wind power is just sufficient to drive away the accreting gas,
ie to $\dot M\sim\sigma^5/\epsilon Gc^2
\sim2\times10^{-3}(\sigma/100{\rm km s}^{-1})^5$~M$_\odot$~yr$^{-1}$.
The hole grows until it has a mass 
$M\sim2\times10^5(\sigma/100{\rm km s}^{-1})^5$~M$_\odot$
after which point the accretion rate will probably be Eddington-limited
and the hole mass will grow exponentially 
until the wind associated with a sub-critical accretion
(formed with an efficiency 
$\epsilon_W$) can limit the accretion again. If we allow this 
phase to continue for a dynamical time, $\sim3r_{{\rm gal}}/\sigma$
then the final black hole mass will be $M\sim3\sigma^4r_{{\rm gal}}/
\epsilon_W Gc^2\sim6\times10^6(\sigma/100{\rm km s}^{-1})^4
(r_{{\rm gal}}/10{\rm kpc})(\epsilon_W/10^{-3})^{-1}$~M$_\odot$.
Choosing plausible values $\epsilon_W\sim10^{-3};r_{{\rm gal}}
\sim10{\rm kpc}$ give a relation similar to that found 
by Ferrarese et al. (2000) and 
Gebhardt et al. (2000), [Marconi, van der Marel].  

Obviously, this is a gross over-simplification of the many complex processes at
work, simultaneously, in a nascent galaxy. Radiative cooling and star formation
will be stimulated in the surrounding gas and this may cause additional heating.
The direct and indirect influence of the hole may soften the potential. 
Rotation and mergers are surely relevant.  
Most importantly, I have 
not specified whether or not the gas falls in towards the nucleus or moves
out towards $r_{{\rm gal}}$ -- whether galaxies form from the outside in or
the inside out. This raises a radical possibility.
Perhaps early-type galaxies are so called because their central black
holes grow quickly and expel enough gas to prevent the subsequent formation of disks.
This scheme, which is subject to
hydrodynamic simulation, has clear, observable implications.
\acknowledgments
I am indebted to many collaborators, including Avery Broderick,
Marco Salvati, Todd Small and especially 
Mitch Begelman for discussion of the above. Support under NSF grant
AST 99-00866 and NASA grant 5-2837 is gratefully acknowledged.


\begin{references}
\reference Agol. E. \& Krolik, J. 2000, \apj, 528, 161
\reference Baganoff, F. et al. 2000 Paper presented at November 2000 HEAD meeting, Hawaii
\reference Blandford, R. D. \& Begelman, M. C. 1999, \mnras, 303, L1 
\reference Ferrarese, L. \& Merritt, D. 2000, \apj, 539, L9
\reference Gebhardt, K. et al. 2000 \apj, 539, L13
\reference Lee, H-K, Wijers, R. A. M. J. \& Brown, G. E. 1999, Phys. Rep., 325, 83
\reference Levinson, A. \& Blandford, R. D. 1995, \mnras, 274, 717
\reference Nandra, P. 2000 Proc.  X-ray Astronomy '99 - Stellar Endpoints, AGN and 
   the Diffuse Background, in press
\reference Natarajan, P. \& Pringle, J. 1998, \apj, 506, L97
\reference Ruffini, R. \& Wilson, J. R. 1975, \prd, 12, 2959
\reference Silk, J. \& Rees, M. 1998, \aap, 331, L1
\reference Wardle, J. F. C. et al. 1998, Nature, 395, 457 
\reference Wehrle, A. et al. 1998, \apj, 497, 178
\reference Wilson, A., Shopbell, P. \& Young, A. 2000, \apj, in press 
\end{references}
\end{document}